\documentclass[runningheads]{llncs}
\usepackage[T1]{fontenc}
\usepackage{amsmath, xcolor}
\usepackage{graphicx,verbatim}
\usepackage{hyperref}
\begin{document}
\title{Physiological neural representation for personalised tracer kinetic parameter estimation from dynamic PET}
\titlerunning{Physiological neural representation for dynamic PET}

\author{Kartikay Tehlan\inst{1,2}\orcidID{0000-0003-2417-1202} \and \\
Thomas Wendler\inst{1,2,3,4}\orcidID{0000-0002-0915-0510}}

\authorrunning{K. Tehlan and T. Wendler}

\institute{Department of diagnostic and interventional Radiology and Neuroradiology, University Hospital Augsburg, Stenglinstr. 2, 86156 Augsburg, Germany \email{kartikay.tehlan@med.uni-augsburg.de} \and 
Computer-Aided Medical Procedures and Augmented Reality, Technical University of Munich, Boltzmannstr. 3, 85748 Garching bei München, Germany \and
Digital Medicine, University Hospital Augsburg, Gutenbergstr. 7, 86356, Neusäß, Germany \and
Center of Advanced Analytics and Predictive Sciences, University of Augsburg, Universitätsstr. 2, 86159 Augsburg, Germany}

\maketitle              %
\begin{abstract}
Dynamic positron emission tomography (PET) with [$^{18}$F]FDG enables non-invasive quantification of glucose metabolism through kinetic analysis, often modelled by the two-tissue compartment model (TCKM). However, voxel-wise kinetic parameter estimation using conventional methods is computationally intensive and limited by spatial resolution. Deep neural networks (DNNs) offer an alternative but require large training datasets and significant computational resources. To address these limitations, we propose a physiological neural representation based on implicit neural representations (INRs) for personalized kinetic parameter estimation. INRs, which learn continuous functions, allow for efficient, high-resolution parametric imaging with reduced data requirements. Our method also integrates anatomical priors from a 3D CT foundation model to enhance robustness and precision in kinetic modelling. We evaluate our approach on an [$^{18}$F]FDG dynamic PET/CT dataset and compare it to state-of-the-art DNNs. Results demonstrate superior spatial resolution, lower mean-squared error, and improved anatomical consistency, particularly in tumour and highly vascularized regions. Our findings highlight the potential of INRs for personalized, data-efficient tracer kinetic modelling, enabling applications in tumour characterization, segmentation, and prognostic assessment. The code is available at: \url{https://github.com/tkartikay/PhysNRPET}


\end{abstract}

\section{Introduction}

Dynamic positron emission tomography (PET) with [$^{18}$F]FDG has allowed for non-invasive assessment of glucose metabolism in tissues through kinetic analysis, in particular with the irreversible two-compartment-kinetic model (TCKM). The kinetic parameters thus obtained, $K_1$, $k_2$, and $k_3$ represent the rates of transport of the tracer between the blood plasma and the cells (influx and efflux) and the rate of phosphorylation of glucose in the cells, respectively. $V_B$, an associated kinetic parameter describes the blood volume fraction, to correct for the partial volume effects arising from the limitations of the scanner spatial resolutions \cite{dimitrakopoulou2021kinetic}. Since glucose metabolism differs for metastatic lesions from non-metastatic tissues with increased uptake, kinetic parameters have been advantageous in successful diagnosis, complementing Standard Uptake Values (SUV) values from static PET images \cite{wumener2022dynamic}. Phosphorylation rate ($k_3$) is significant metric for tumour differentiation in colorectal cancers \cite{strauss2007assessment} and breast cancers \cite{kajary2020dynamic}, and the perfusion parameter ($K_1$) can be a marker of angiogenesis in breast cancers \cite{cochet2012evaluation}. Kinetic parameters, carrying more information than SUV, can play a vital role as additional biomarkers for tumour heterogeneity analysis, precise segmentations, and prognostic predictions.

Conventional methods based on non-linear least squares regression used to perform kinetic analysis on volumes of interest lack the spatial neighbourhood understanding and are slowed down when applied to derive voxel-based parametric images. These challenges can be overcome through the use of deep neural networks (DNNs) to derive the parametric images \cite{cui2022unsupervised,huang2022parametric,liang2023combining}. Self-supervised DNNs have recently been proposed to generate the $K_1, k_2, k_3, V_B$ parametric images which satisfy the TCKM \cite{de2023self}. These methods, in turn, require large amount of training data to achieve accurate results, and while they are quick to predict the outputs once trained, the training of these networks itself is a slow and memory-intensive process, balanced by reducing the resolution of the training dataset, and slicing it into 2D+t data instead of processing the entire 3D+t data at once.

Implicit neural representations (INRs) are neural networks that learn continuous functions, which can parametrise signals, such as mapping continuously the image intensities from the input spatial coordinates. With steady adoptions in the medical imaging field due to their memory efficiency, facilitation of signal reconstructions, interpolation capabilities, and personalisation \cite{molaei2023implicit}, INRs have the potential to overcome the data limitation issues faced by DNNs in generation of parametric images in dynamic PET modelling. Since INRs can be optimised on single patient acquisitions, they can be adapted to individual patient anatomy and physiology \cite{shi2024implicit}, leading to personalised neural networks following the paradigm of precision medicine.

Encoding the dynamic physiological processes in the INR also allows their integration with downstream models that are trained for classification of malignancies, or early detection of metastases, etc., as well as with foundation models that extract features from the associated anatomical data such as CT scans, thereby enabling feature-rich end-to-end pipelines for aiding the clinical routines.

Thus, guided by their promise, we propose the use of INRs for modelling dynamic PET to generate kinetic parameters. Our contribution is:

\begin{itemize}
    \item the introduction of a physiological neural representation for tracer kinetic modelling,
    \item its extension by integrating anatomical information and features extracted from corresponding CT scan using a 3D CT foundation model, and
   \item its evaluation using an [$^{18}$F]FDG dynamic PET/CT dataset using comparative analysis against state-of-the-art DNNs in terms of temporal and spatial resolution, as well as goodness of fit.
\end{itemize}

\subsection{Related Works}

Voxel-wise estimation of kinetic parameters have been explored by solving the non-linear optimisation problem of fitting the model-predicted curves onto the measured data \cite{besson202018,wang2022total}. The data in these studies was limited to smallest frames of 10s resolution. With the advancements in long axial field of view scanners, and the availability of improved temporal resolution, the kinetic parameters have also been estimated for increased resolution of 2s frames \cite{sari2022first}. Sari et al. provide reference values for kinetic parameters for volumes of interest in oncological patients, and have also shown the voxel-wise generation of TCKM. Building upon this work, de Benetti el al. proposed DNNs for the generation of parametric images from the dynamic PET data, relying on self-supervised training of the DNN through the integration of the kinetic model into the loss function \cite{de2023self}. To parametrise tracer kinetics through encoding the PET reconstruction and one compartment model in neural networks, a hybrid physics- and data-driven approaches is proposed to incorporate the mathematical expressions defining the kinetics \cite{ye2024hybrid}. Shao et al. propose employing DNNs to discretise the data, and solving the ODEs that define the kinetic model, where the tracer concentration in each compartment is represented with separate fully connected DNNs, and the kinetic parameters are optimised while minimising the ODE loss \cite{shao2023fast}.

Since INRs are based on the assumption of continuity as an \textit{a priori} \cite{molaei2023implicit}, these representations can be made robust to noise and further constrained anatomically through image-derived priors, such as extracted from large-scale "foundation" models. Large-scale models trained on high-resolution oncological data can provide knowledge of vascular architecture, tissue heterogeneity, and image-based biomarkers as embeddings or input features to the task-specific models. Neural networks that encode the tracer kinetics could similarly benefit from the anatomical understanding of macroscopic blood supply routes, sharp gradients marking organ bounds,  regions of tissue similarities and anomalies, etc. A foundation model trained on high-resolution 3D CT oncological datasets with understanding of underlying biology, and shown to improve performance of downstream tasks, particularly with limited dataset, and resilient to input variation has been present by Pai et al \cite{pai2024foundation}.

\section{Materials and Methods}

\subsection{Dataset}

The dataset consists of 24 patients with various tumour types undergoing dynamic PET imaging (average injected activity: 235 ± 51 MBq of [$^{18}$F]FDG), immediately followed by a CT scan using a Siemens Biograph Vision Quadra long-field-of-view PET/CT device. The PET acquisition lasted 65 minutes for all patients and consisted of 62 frames with varying durations: 2 seconds for the initial frames, 30 seconds after the first 2 minutes, and up to 5 minutes per frame toward the end of the acquisition. PET images were reconstructed with a voxel size of 1.65 × 1.65 × 1.65 mm³ and subsequently filtered using a 2 mm full-width at half maximum (FWHM) Gaussian filter to generate \textbf{high-resolution images (HiRes)}. To match the resolution used by De Benetti et al., we downsampled the PET images to 2.5 × 2.5 × 2.5 mm³, producing \textbf{low-resolution images (LoRes)}. The CT scans were acquired in low-dose mode (voltage: 120 kV, tube current: 25 mA) and reconstructed with a voxel size matching that of the high-resolution PET images.

\subsection{Preprocessing and Normalisation of Data and Input}

Dynamic PET data were first normalised to the range $[0,1]$ by dividing by the maximum of the image-derived input function, which was on the order of $\sim 200\;\mathrm{kBq/ml}$. The corresponding CT data were similarly rescaled to the interval $[0,1]$, where $-1024\;\mathrm{H.U.}$ was mapped to 0 and $2048\;\mathrm{H.U.}$ was mapped to 1.

Using the normalised CT volume, the CT foundation model \cite{pai2024foundation} extracted 4096 features for each voxel, by taking as input $50\times 50 \times 50$ neighbourhood centred on that voxel. These features served as an optional input to the network, depending on the variant of the INR under consideration.

Spatial coordinates $x \in \mathrm{R^3}$ were first normalised to the interval $[-1,1]$. Each normalised coordinate was encoded using Gaussian Fourier Features (GFF) \cite{tancik2020fourier} \cite{rahimi_2007} drawn from a zero-mean Gaussian distribution, with a standard deviation chosen to control frequency bandwidth. A total of 256 frequencies were sampled, and a multiplier of 10 was applied to spread the frequency spectrum. Specifically, for each coordinate $\Vec{x}$, the encoding was formed by computing:
$\gamma(\Vec{x}) = [sin(\Vec{2\pi Bx}), cos(\Vec{2\pi Bx})]^T$
where $\Vec{B}$ is a randomly generated matrix of size $\Vec{256\times 3}$. The choice of 256 frequencies and a multiplier of 10 followed typical recommendations in the literature and was not varied \cite{tancik2020fourier}. In all cases, the spatial domain (2D or 3D) was sampled, and the resulting coordinates were normalised to the interval $[-1,1]$ prior to being processed with GFF. Where relevant (CTHU variant, see below), the voxel-wise H.U. value, also normalised to $[0,1]$, was appended to the normalised coordinates to form an extended input vector. This augmented input was then encoded with GFF. However, in the variant utilising the foundation model features (CTFM variant, see below), the 4096-dimensional feature vector was incorporated directly, without GFF encoding, while GFF was still applied to the spatial coordinates. 

The encoded features were passed to a SIREN \cite{sitzmann2020implicit} network with three hidden layers. Each hidden layer comprised 512 units and used sine activations, following the standard weight-initialisation scheme described in the SIREN work to preserve gradient stability. The final layer was a linear layer producing four outputs: $K_1$, $k_2$, $k_3$, and $V_b$. These outputs correspond to the parameters in the 2TC kinetic model governed by a system of ordinary differential equations (ODEs). Maintaining differentiability from inputs to outputs facilitated direct gradient-based training in conjunction with the ODE formulation.

Parameter estimation was performed by minimising the mean-squared error (MSE) between the time activity curve (TAC) generated from the predicted parameters $(K_1, k_2, k_3, V_b)$ and real TAC values measured at the input coordinate in the dynamic PET. Training used the Adam optimiser with a learning rate of $1e^{-5}$. The network was trained for 100 epochs and all hyperparameters, including the number of frequencies, the sine activation design, and the network depth, were adapted from SIREN and GFF examples without additional ablation experiments\cite{byra2023exploring,reed2021dynamic,shen2022nerp}.

\subsection{Loss function}

Using the estimated kinetic parameters $\bigl(\hat{k}_1, \hat{k}_2, \hat{k}_3, \hat{v}_b\bigr)$ from the INR for each voxel coordinate, the corresponding TAC was generated by solving the TCKM ODEs. The predicted TAC was then compared against the observed TAC at the discrete time points where measurements were available. This comparison employed an MSE loss, defined as:

\begin{equation}
    \text{MSE} \;=\; \frac{1}{N}\sum_{i=1}^N \bigl(\widehat{\text{TAC}}(t_i) \;-\; \text{TAC}(t_i)\bigr)^2,
\end{equation}

where $\widehat{\text{TAC}}(t_i)$ denotes the predicted TAC value at time $t_i$, and $\text{TAC}(t_i)$ is the corresponding measured value. N is the number of observations. The mean and standard deviation of these reconstruction errors were computed across the full 3D spatial domain with time (3D+t), selected 2D slices with time (2D+t), and at each individual voxel with time (voxel+t). These metrics were used to compare the performance of the proposed INR-based predictions against those from a conventional neural network approach, providing a quantitative measure of accuracy in recovering voxel-wise kinetic parameters.

\section{Results}

We evaluated the proposed architecture in several variants: HiRes and LoRes, 2D (single slice) and 3D (a stack of 10 adjacent slices) - resulting in the variants inr-HiRes-2D, inr-LoRes-2D, inr-HiRes-3D and inr-LoRes-3D. Additionally, we tested configurations incorporating the Hounsfield unit (HU) values from the CT scan (only for HiRes and in 2D - variant CTHU-2D) and those using features extracted with the foundation model (variant CTFM-2D). We also evaluated the neural network proposed by De Benetti et al. in both 2D (single-slice) and 3D (stack of 10 adjacent slices) configurations (2D and 3D baselines) \cite{de2023self}.

As comparison metrics, we used MSE with its standard deviation (SD), as well as training time, inference time, and memory utilization. All experiments were conducted on a MacBook with an M3 Pro (12-core CPU, Metal 3, 36GB RAM). The results of this evaluation are summarized in Tab. \ref{tab:results}.

\begin{table}[!h]
\centering
\caption{Quantitative evaluation of INR variants and comparison with the spatio-temporal neural network of De Benetti et al.}\label{tab:results}
\begin{tabular}{|l|l|l|l|l|l|}
\hline
Model & Voxel  & MSE($ \pm $SD) & Training  & Inference  & Memory  \\
 & Resolution & & Time & Time & Utilisation \\
\hline
\hline
FDB112-3D & 2.5 mm & 0.066 $\pm$ 0.255 & - & 62 s & - \\
INR-LoRes-3D & 2.5 mm & 0.009 $\pm$ 0.093 & 24.3 min & 66 s & 6.7 GB\\
\hline
FDB112-2D & 2.5 mm & 0.073 $\pm$ 0.268 & - & 0.15 s * & - \\
INR-LoRes-2D & 2.5 mm & 0.009 $\pm$ 0.093 & 24.1 min & 6 s & 4.3 GB \\
\hline
INR-2D-HiRes & 1.65 mm & 0.009 $\pm$ 0.094 & 25.2 min & 8 s & 11.9 GB \\
HU-INR-2D-HiRes & 1.65 mm & 0.009 $\pm$ 0.094 & 25.4 min & 8 s & 12.4 GB\\
FM-INR-2D-HiRes & 1.65 mm & 0.009 $\pm$ 0.094 & 25.7 min & 9 s & 13 GB \\
\hline
\end{tabular}
\end{table}

\begin{figure}
\centering
\includegraphics[width=\textwidth]{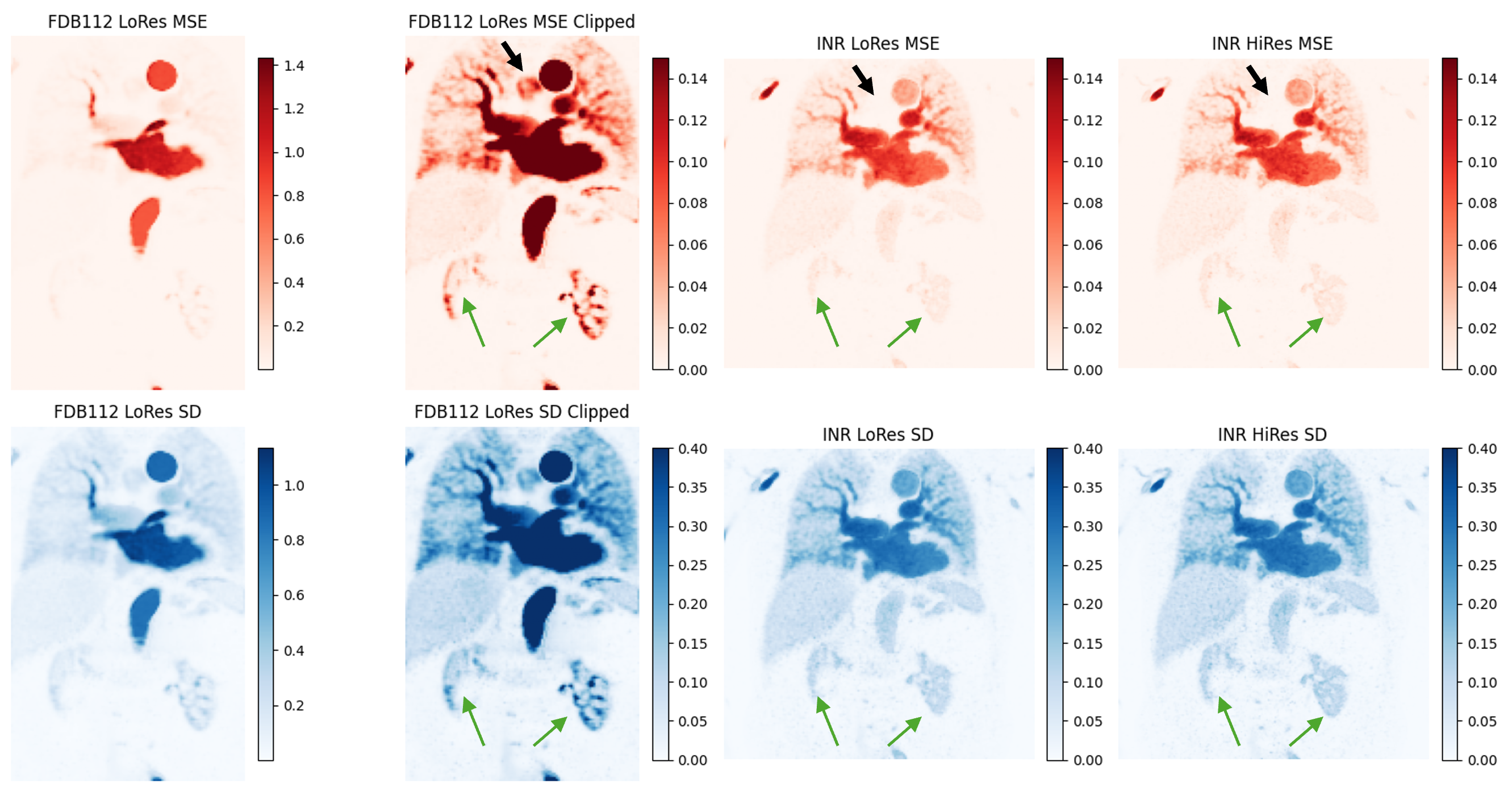}
\caption{MSE (top row, red) and SD (lower row, blue) for a coronal slice of an exemplary patient. MMSE is higher on average over the complete slice (see FDB112 LoRes MSE Clipped vs. INR LoRes MSE), but particularly in the location of tumour (black arrow) and kidneys (green arrow), ignoring the blood pool (visible larger vessels). Also SD images show less variation on the MSE over time for our model. The comparison of LoRes and HiRes variants of our model (righ two columns) shows a good degree of agreement with more details in the HiRes image.}
\label{fig:MSE_coronal}
\end{figure}

\begin{figure}
\centering
\includegraphics[width=\textwidth]{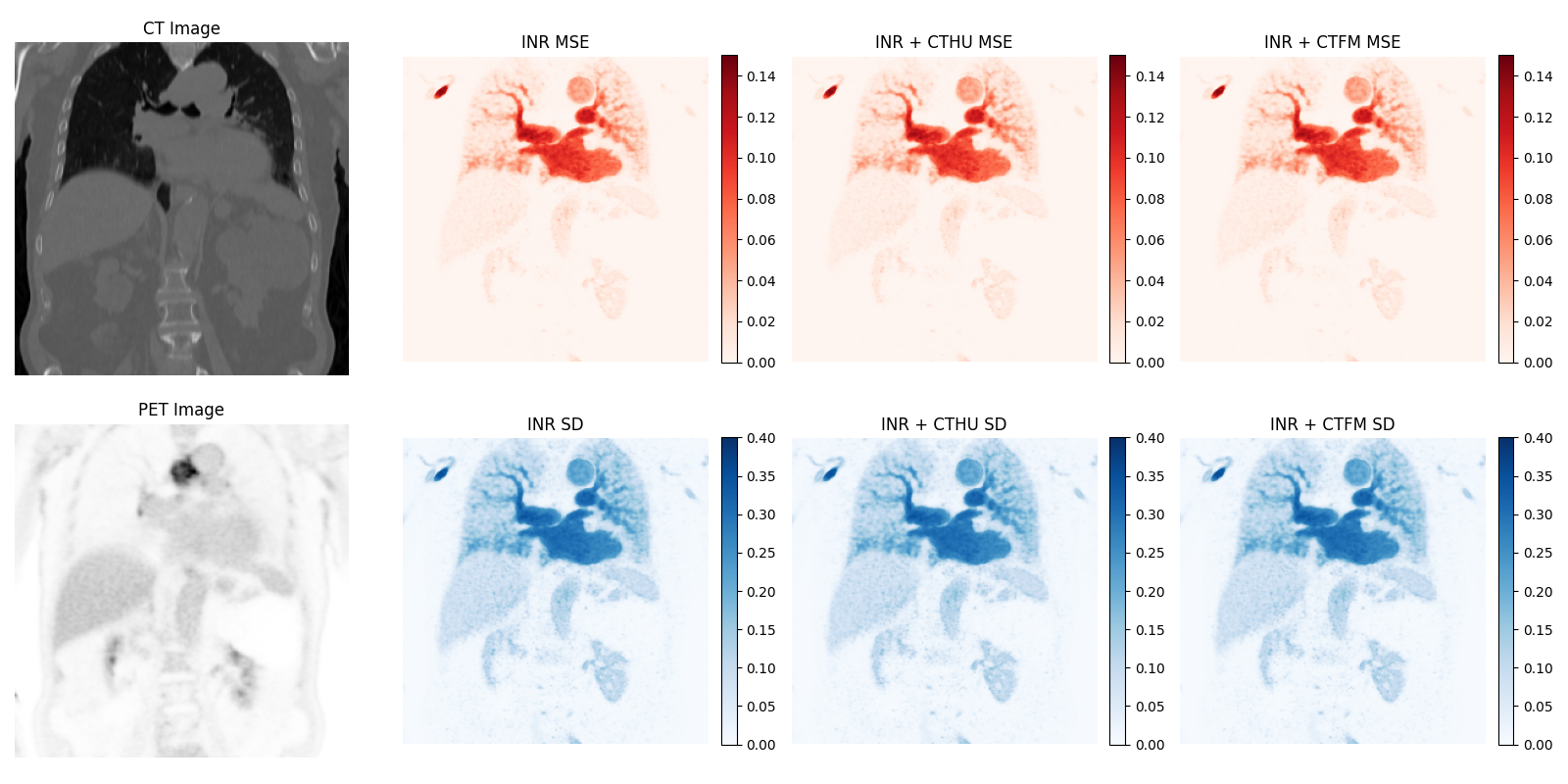}
\caption{Comparison of different variants of HiRes models in a coronal slice of an exemplary patient. The first column shows the corresponding CT and  static PET images. With increased input information, the degree of granularity slightly increases while MSE average and SD range remaining almost identical.} \label{fig:MSE_coronal_2}
\end{figure}

\begin{figure}
\centering
\includegraphics[width=\textwidth]{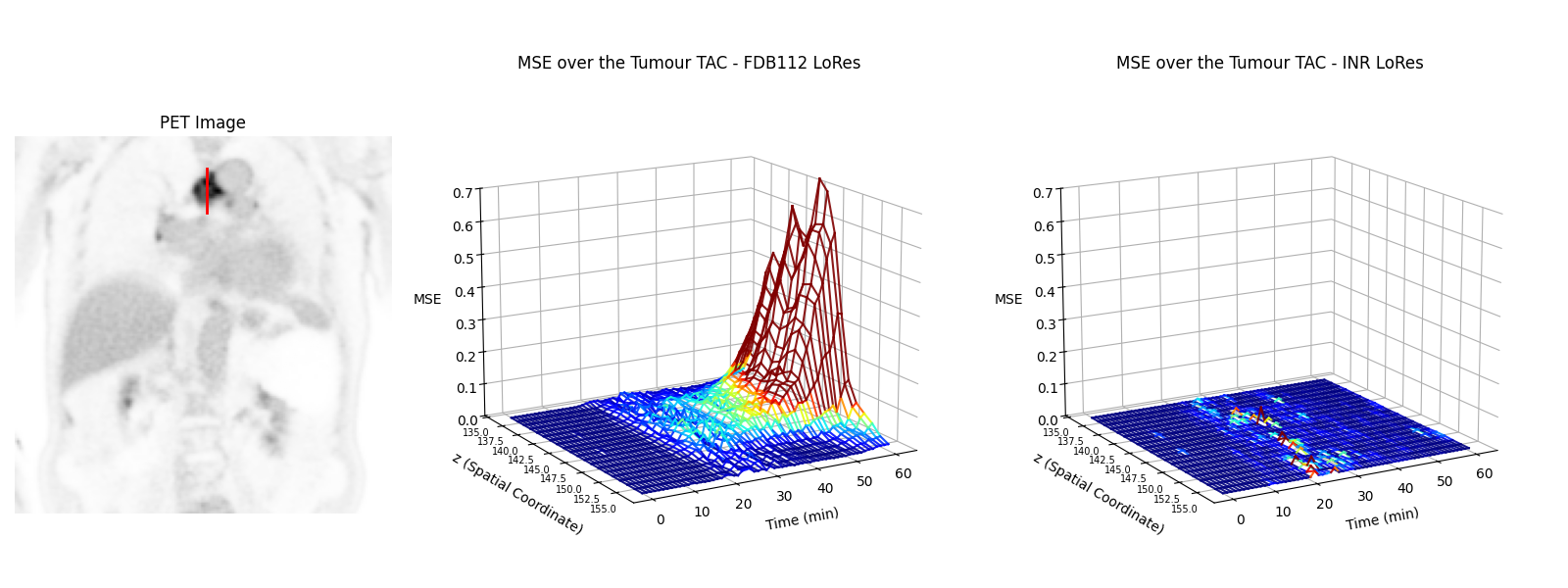}
\caption{MSE along segment crossing a mediastinal lymph node metastasis (depicted in corresponding PET slice, left). The plots show the MSE as a function of space (axis going from left to bottom of the graphs) and time (axis going from bottom to right). The center plot is the MSE for the 2D baseline, while the right plot shows the inr-LoRes-2D variant. While the baseline shows the error in the center of the lesion at a late time point of the acquisition, the INR shows significantly lower MSE and rather placed at the peak of the IDIF - not visible in the graph.} \label{fig:tumorTAC}
\end{figure}

To qualitatively visualize the impact of the different INR variants, we visualize the mean MSE and its SD in Figs. \ref{fig:MSE_coronal} and \ref{fig:MSE_coronal_2} for an exemplary patient. This gives an idea of which organs the measured and the calculated TACs depart in average (mean MSE), and how large the spread of the error (SD) is. Additionally, Fig. \ref{fig:tumorTAC} %
shows for a particular segment across the tumour or the left kidney, respectively when in time and space the MSE is highest.

\begin{figure}
\centering
\includegraphics[width=\textwidth]{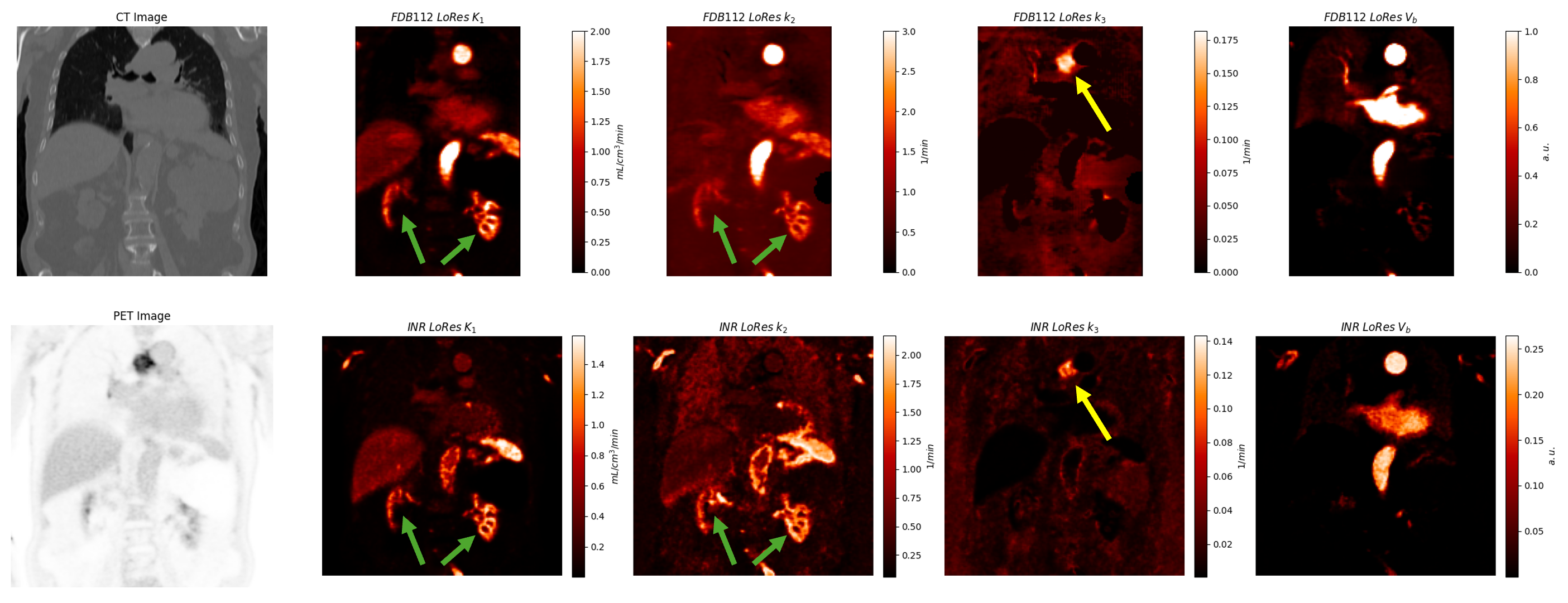}
\caption{Coronal slice of parametric images generated with the 2D baseline (upper row second to last columns) and with the inr-LoRes-2D variant. The INR parameters show signficantly higher resolution in terms of more details and sharper edges. In concrete, the kidneys (green arrow, best visible in $K_1$ and $k_2$ images) and tumour (yellow arrow, best seen in $k_3$ image) have more clearer boundaries, capturing their heterogeneity. The tumour also shows a higher contrast to other structures pointing at higher specificity.}
\label{fig:Ki}
\end{figure}

\section{Discussion}

This study applies implicit neural representations (INRs) to tracer kinetic modelling in dynamic [$^{18}$F]FDG PET. Unlike non-linear least squares regression or large-scale DNNs, INRs approximate the continuous spatio-temporal function of patient-specific PET signals. Such a representation requires fewer training examples than conventional DNNs, making it useful when data availability is limited. In addition, modelling kinetics through an INR allows more flexibility in handling anatomical and physiological variability across patients.

We integrated anatomical information from CT by incorporating HU values and, in a separate variant, 3D CT foundation model features. Both approaches provided comparable MSE but showed slightly faster convergence (not shown), suggesting that anatomical descriptors can guide the estimation of kinetic parameters. The proposed INRs consistently achieved lower MSE than a baseline DNN, likely due to their ability to capture spatio-temporal dynamics via a continuous, data-efficient mapping.

Performance across INR variants was similar in accuracy, though convergence speed and memory demands differed. Foundation-model-based INRs converged faster but required more memory. Nonetheless, LoRes variants of our INR reduced the resource footprint, enabling training on standard hardware setups. %

Comparisons with a baseline DNN proved that INRs show lower errors, especially in tumour and kidney regions, which often exhibit rapid or heterogeneous tracer kinetics. This highlights the robust parameter estimation of INRs when organ- or lesion-specific kinetic variability must be captured.

Accurate kinetic parameters ($K_1$, $k_2$, $k_3$, $V_b$) can reveal metabolic properties that aid in tumour assessment and therapy evaluation. The continuous form of INRs and their integration with anatomical descriptors can potentially reduce segmentation errors and support downstream tasks such as lesion classification. Further research could expand the scope by incorporating additional imaging modalities, testing on larger datasets, or refining network architectures and Fourier feature encodings. This future work may improve accuracy, promote generalisability, and further lower the hardware demands for advanced tracer kinetic modelling.

\newpage
\bibliography{Main.bib}

\begin{thebibliography}{10}
\providecommand{\url}[1]{\texttt{#1}}
\providecommand{\urlprefix}{URL }
\providecommand{\doi}[1]{https://doi.org/#1}

\bibitem{besson202018}
Besson, F.L., Fernandez, B., Faure, S., Mercier, O., Seferian, A., Mignard, X., Mussot, S., Le~Pechoux, C., Caramella, C., Botticella, A., et~al.: 18 f-fdg pet and dce kinetic modeling and their correlations in primary nsclc: first voxel-wise correlative analysis of human simultaneous [18f] fdg pet-mri data. EJNMMI research  \textbf{10},  1--13 (2020)

\bibitem{byra2023exploring}
Byra, M., Poon, C., Rachmadi, M.F., Schlachter, M., Skibbe, H.: Exploring the performance of implicit neural representations for brain image registration. Scientific Reports  \textbf{13}(1),  17334 (2023)

\bibitem{cochet2012evaluation}
Cochet, A., Pigeonnat, S., Khoury, B., Vrigneaud, J.M., Touzery, C., Berriolo-Riedinger, A., Dygai-Cochet, I., Toubeau, M., Humbert, O., Coudert, B., et~al.: Evaluation of breast tumor blood flow with dynamic first-pass 18f-fdg pet/ct: comparison with angiogenesis markers and prognostic factors. Journal of Nuclear Medicine  \textbf{53}(4),  512--520 (2012)

\bibitem{cui2022unsupervised}
Cui, J., Gong, K., Guo, N., Kim, K., Liu, H., Li, Q.: Unsupervised pet logan parametric image estimation using conditional deep image prior. Medical image analysis  \textbf{80},  102519 (2022)

\bibitem{de2023self}
De~Benetti, F., Simson, W., Paschali, M., Sari, H., Rominger, A., Shi, K., Navab, N., Wendler, T.: Self-supervised learning for physiologically-based pharmacokinetic modeling in dynamic pet. In: International Conference on Medical Image Computing and Computer-Assisted Intervention. pp. 290--299. Springer (2023)

\bibitem{dimitrakopoulou2021kinetic}
Dimitrakopoulou-Strauss, A., Pan, L., Sachpekidis, C.: Kinetic modeling and parametric imaging with dynamic pet for oncological applications: general considerations, current clinical applications, and future perspectives. European journal of nuclear medicine and molecular imaging  \textbf{48},  21--39 (2021)

\bibitem{huang2022parametric}
Huang, Z., Wu, Y., Fu, F., Meng, N., Gu, F., Wu, Q., Zhou, Y., Yang, Y., Liu, X., Zheng, H., et~al.: Parametric image generation with the uexplorer total-body pet/ct system through deep learning. European Journal of Nuclear Medicine and Molecular Imaging  \textbf{49}(8),  2482--2492 (2022)

\bibitem{kajary2020dynamic}
Kaj{\'a}ry, K., Lengyel, Z., T{\H{o}}k{\'e}s, A.M., Kulka, J., Dank, M., T{\H{o}}k{\'e}s, T.: Dynamic fdg-pet/ct in the initial staging of primary breast cancer: clinicopathological correlations. Pathology \& Oncology Research  \textbf{26},  997--1006 (2020)

\bibitem{liang2023combining}
Liang, G., Zhou, J., Chen, Z., Wan, L., Wumener, X., Zhang, Y., Liang, D., Liang, Y., Hu, Z.: Combining deep learning with a kinetic model to predict dynamic pet images and generate parametric images. EJNMMI physics  \textbf{10}(1), ~67 (2023)

\bibitem{molaei2023implicit}
Molaei, A., Aminimehr, A., Tavakoli, A., Kazerouni, A., Azad, B., Azad, R., Merhof, D.: Implicit neural representation in medical imaging: A comparative survey. In: Proceedings of the IEEE/CVF International Conference on Computer Vision. pp. 2381--2391 (2023)

\bibitem{pai2024foundation}
Pai, S., Bontempi, D., Hadzic, I., Prudente, V., Soka{\v{c}}, M., Chaunzwa, T.L., Bernatz, S., Hosny, A., Mak, R.H., Birkbak, N.J., et~al.: Foundation model for cancer imaging biomarkers. Nature machine intelligence  \textbf{6}(3),  354--367 (2024)

\bibitem{rahimi_2007}
Rahimi, A., Recht, B.: Random features for large-scale kernel machines. In: Platt, J., Koller, D., Singer, Y., Roweis, S. (eds.) Advances in Neural Information Processing Systems. vol.~20. Curran Associates, Inc. (2007)

\bibitem{reed2021dynamic}
Reed, A.W., Kim, H., Anirudh, R., Mohan, K.A., Champley, K., Kang, J., Jayasuriya, S.: Dynamic ct reconstruction from limited views with implicit neural representations and parametric motion fields. In: Proceedings of the IEEE/CVF International Conference on Computer Vision. pp. 2258--2268 (2021)

\bibitem{sari2022first}
Sari, H., Mingels, C., Alberts, I., Hu, J., Buesser, D., Shah, V., Schepers, R., Caluori, P., Panin, V., Conti, M., et~al.: First results on kinetic modelling and parametric imaging of dynamic 18 f-fdg datasets from a long axial fov pet scanner in oncological patients. European journal of nuclear medicine and molecular imaging pp. 1--13 (2022)

\bibitem{shao2023fast}
Shao, W., Chen, Y., Li, N., Yang, Z., Meng, X., Xie, Z.: Fast and robust estimation of kinetic parameters in dynamic pet imaging using neural network-based discretization method (2023)

\bibitem{shen2022nerp}
Shen, L., Pauly, J., Xing, L.: Nerp: implicit neural representation learning with prior embedding for sparsely sampled image reconstruction. IEEE Transactions on Neural Networks and Learning Systems  \textbf{35}(1),  770--782 (2022)

\bibitem{shi2024implicit}
Shi, J., Zhu, J., Pelt, D.M., Batenburg, K.J., Blaschko, M.B.: Implicit neural representations for robust joint sparse-view ct reconstruction. arXiv preprint arXiv:2405.02509  (2024)

\bibitem{sitzmann2020implicit}
Sitzmann, V., Martel, J., Bergman, A., Lindell, D., Wetzstein, G.: Implicit neural representations with periodic activation functions. Advances in neural information processing systems  \textbf{33},  7462--7473 (2020)

\bibitem{strauss2007assessment}
Strauss, L.G., Klippel, S., Pan, L., Sch{\"o}nleben, K., Haberkorn, U., Dimitrakopoulou-Strauss, A.: Assessment of quantitative fdg pet data in primary colorectal tumours: which parameters are important with respect to tumour detection? European journal of nuclear medicine and molecular imaging  \textbf{34},  868--877 (2007)

\bibitem{tancik2020fourier}
Tancik, M., Srinivasan, P., Mildenhall, B., Fridovich-Keil, S., Raghavan, N., Singhal, U., Ramamoorthi, R., Barron, J., Ng, R.: Fourier features let networks learn high frequency functions in low dimensional domains. Advances in neural information processing systems  \textbf{33},  7537--7547 (2020)

\bibitem{wang2022total}
Wang, G., Nardo, L., Parikh, M., Abdelhafez, Y.G., Li, E., Spencer, B.A., Qi, J., Jones, T., Cherry, S.R., Badawi, R.D.: Total-body pet multiparametric imaging of cancer using a voxelwise strategy of compartmental modeling. Journal of Nuclear Medicine  \textbf{63}(8),  1274--1281 (2022)

\bibitem{wumener2022dynamic}
Wumener, X., Zhang, Y., Wang, Z., Zhang, M., Zang, Z., Huang, B., Liu, M., Huang, S., Huang, Y., Wang, P., et~al.: Dynamic fdg-pet imaging for differentiating metastatic from non-metastatic lymph nodes of lung cancer. Frontiers in oncology  \textbf{12},  1005924 (2022)

\bibitem{ye2024hybrid}
Ye, Y., Liu, H., Wang, L.: Hybrid kinetics embedding framework for dynamic pet reconstruction. arXiv preprint arXiv:2403.07364  (2024)

\end{thebibliography}

\end{document}